\documentclass[10pt]{article}
\usepackage{amssymb}
\usepackage{amsthm}
\usepackage{authblk}
\usepackage{setspace}
\usepackage{hyperref}
\usepackage[textwidth=6.5in,textheight=8.25in,centering]{geometry}
\usepackage[pdftex]{graphicx}

\onehalfspace
\begin{document}
\begin{center}
{\LARGE{\textbf{Tent--Shaped Surface Morphologies of Silicon:   Texturization by Metal Induced Etching}}}

\vspace{0.5 cm}
\textit{Deepika Poonia $ ^{\$} $, Priyanka Yogi $^{\$}$, Suryakant Mishra, Shailendra K. Saxena,  Swarup Roy, Pankaj R. Sagdeo and Rajesh Kumar}\footnote{Corresponding author email: rajeshkumar@iiti.ac.in} \footnote{\href{http://magse.webs.com/} {\textbf{http://magse.webs.com/}}}

\vspace{0.5 cm}
Material Research Laboratory, Discipline of Physics \& MEMS, Indian Institute of Technology Indore, Simrol-453552, Madhya Pradesh, India

$^\$ $ \textit{Authors having equal contribution}

\vspace{1 cm}
ABSTRACT

\end{center}
 Nano--metal/semiconductor junction dependent porosification of silicon (Si) has been studied here. The silicon (Si) nanostructures (NS) have been textured on n-- and p-- type silicon wafers using Ag and Au metal nano particles induced chemical etching. The combinations of n--Si/Ag and p--Si/Au form ohmic contact and result in the same texturization on the Si surface on porosification  where tent--shaped morphology has been observed consistently with n-- and p--type Si. Whereas, porosification result in different surface texturization for other two combinations (p--Si/Ag and n--Si/Au) where Schottkey contacts are formed. Quantitative analysis have been done using ImageJ to process the  SEM images of SiNS, which confirms that the tent like SiNS are formed when etching of silicon wafer is done by AgNPs and AuNPs on n and p type Si wafer respectively. These easily prepared sharp tent--shaped Si NSs can be used for enhanced field emission applications.
\vspace{0.5cm}


\section{Introduction}

 Texturization of semiconductors has become one of the great interest of research due to its many applications in various fields[1,2] especially the ones which improves the light trapping efficiency of semiconductors[3]. Teturization of crystalline silicon (c--Si) plays an important role in the fabrication and improvement in efficiency of solar cells[3]. By the texturization of c--Si different shaped and size of Si is achieved by many groups[3–-5]. As the size of Si approaches order of its Bohr radius, various properties vary as compared to its bulk counterpart. Porous materials[6–-8] (and their nanostructures (NSs)) including silicon (Si) has become the great interest of research topic due to its unique properties[9–-13] and many applications[12,14–-16] such as optoelectronics[17-–20], nanoelctronics[21,22], bio--sensors[23,24] as well as energy storage[25] devices and many more. One of the reasons enabling SiNS to be used in these applications is quantum confinement effect. The quantum confinement effect provides the flexibility to tune the properties such as optical, electrical, chemical and thermal properties of Si. The one dimension system give a platform to explore the physical phenomenon happening at nanoscale[13,26,27]  as well as to see the size and dimensionality  dependence of their properties for fundamental research[13,27,28]
 
 Many growth techniques are available for synthesis of SiNS[29-–31], among which metal induced etching (MIE) is one of the simplest and low cost method to fabricate the semiconductor nano structures. MIE, actually is a method of porosification where metal nanoparticles are deposited onto semiconductors followed by etching in an etchant solution as has been discussed in the previous reports[4]. Different mechanisms have been proposed to understand the texturing process in semiconductors by analyzing the morphological modifications during porosification. Various paparemeters affecting the observed surface morphology and end nanostructure form (nanowire, interconnected pores etc) include resistivity of wafer, doping concentration, type of wafer (whether p-- or n--type)[32]. In this connection, role of metal nanoparticles has also been investigated[4]. While investigations are going on it is very clear that at the end it is a chemical etching process and thus involves holes to initiate the etching prosess which needs to reach the wafer surface. Thus the role of metal (nanoparticles) /semiconductor junction can not be neglected as either of the two junctions (Ohmic and Schottkey) can be formed and flow of hole can be affected.
 
 The aim of this present paper is to understand the effect of nano--metal/Si junction on the final surface morphology as a result of porosification by MIE. A tent--shaped morphology of SiNS has been observed when nano--metal/Si junction is Ohmic. In depth study have been carried out by choosing n-- and p--type Si wafer for porosification with gold (Au) and silver (Ag) nanoparticles as catalysts. The non--Ohmic (Schottkey) combinations of nanometal/Si wafers did not result in tent--shaped surface morphologies.  To understand the sole role of nanometal/Si junction on the resultant morphologies other parameters like resistivity, metal deposition time and etching times were kept constant. In order to understand the texturing of Si wafer scanning electron microscopy (SEM) has been carried out.  Quantitative morphological analysis has been done for clear understanding using ImageJ software. 
 
\section{Experimental Details}
Four Si NSs samples have been synthesised to see the effect of metal/Si junction on the porosification during MIE[33] as per the details given in Table--1. Commercially available n-- and p--type Si wafers with resistivity 0.001 $\Omega$--cm have been used for deposition of Ag and Au nanoparticles followed by etching by MIE technique. Cleaned Si wafers of both types were immersed in two different 4.8 M HF solutions containing 5mM AgNO$_3$ and KAuCl$_4$ for deposition of AgNPs and AuNPs for 60 seconds respectively. The four samples (two each with each metal nanoparticles)  were then  transferred to an etching solution containing 4.8M HF \& 0.5M H$_2$O$_2$ for the fabrication of SiNS. After porosification, samples were finished as described elsewhere [4,26,34]. The surface and x--sectional morphologies were recorded using SEM, Supra55 Zeiss. ImageJ software is used to process the SEM images.
\begin{table}
\caption{Details of SiNS  prepared by MIE which are used in the existing study. Resistivity of Si wafer was 0.001 $\Omega$ cm and etching time was fixed at 60 minute for all the samples}
\begin{tabular}{  |c | c | c | c | }
  \hline			
  Sample name &
	Wafer type &	Ag deposition time  (min) &	Au deposition time  (min)
\\
  \hline
 $SN_{Ag}$	& n--type &	1 &	0\\
$SP_{Ag}$	& p--type	& 1	& 0\\
$SN_{Au}$	& n--type &	0	& 1\\
$SP_{Au}$	& p--type &	0 & 	1\\
  \hline  
\end{tabular}

\end{table}

\section{Results and Discussion}
Figure 1 shows the SEM images of all the four samples used in the present study prepared by MIE technique. The top panel (Figure 1) shows the SEM images of SiNS prepared by both n-- and p-- type Si wafer fabricated by AgNPs catalyst and the bottom panel shows the morphology of samples prepared using AuNPs catalyst. From Figure1 it is evident that porosification has successfully taken place on all the samples resulting in porous Si.In case of sample $SN_{Ag}$ some tent like structures is apparent in the top view which can only be confirmed by looking at the tilted view in SEM as will be discussed later. On the other hand,  sample $SP_{Ag}$ the pore like structures are apparent. As it is well known that though the contact between Ag and n--type silicon wafer is different from the contact between Ag and p--type due to that the surface morphology of both n-- and p-- type SiNS may be different. To further confirm that the contact between metal nano particles gives the different texurization of Si wafer another set of sample using AuNPs have been syudied. The bottom panel of Figure 1 shows the SEM images of SiNS prepared using AuNPs. From sample $SN_{Au}$ and $SP_{Au}$ the SiNS have been
\begin{figure}
\begin{center}
\includegraphics[width=12cm]{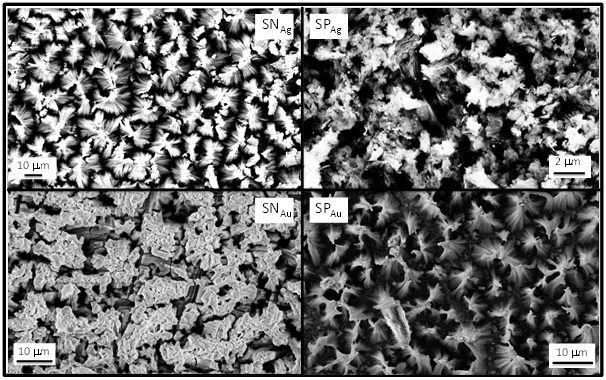}
\caption{The top view SEM images are (surface morphologies) of all the samples mentioned in Table 1.}
\end{center}
\end{figure}
 successfully fabricated throughout the Si wafer of both type but the surface morphology in theses samples of n-- and p-- type is different from each other due to the same contact effect in n and p type Si wafer respectively. Samples $SN_{Ag}$ and $SP_{Au}$ have similar surface morphology and that is of tent--shaped. Another similarity between the two sample is the presence of Ohmic junction between the nano--metal and Si which was used for porosification that resulted in such morphology. It is evident from surface morphologies (Figure 1) that the SiNS morphology  of samples $SN_{Ag}$ and $SP_{Au}$  are similar and are different from samples  sample $SP_{Ag}$ and $SN_{Au}$ which are obtained after porosification of samples with non--Ohmic contact. In order to analyze the morphologies showing tent–like \& porous structures ImageJ analysis of the top view SEM images has been carried out. Surface plots of SEM images in Figure 1 have been shown in Figure 2 along with the corresponding SEM images. The 3D representation in inset of Figure 2 shows the surface plot showing clear tent--shaped morphology. ImageJ analysis of other two samples does not show any tent--shaped structure on the surface.

\begin{figure}
\begin{center}
\includegraphics[width=14cm]{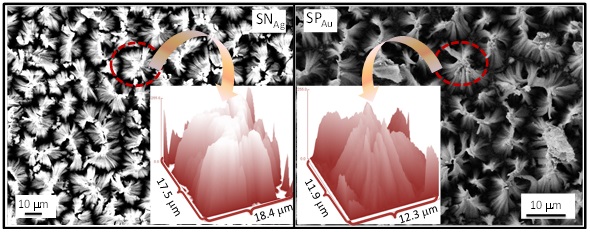}
\caption{SEM images used for the ImageJ to analyse the tent structure in sample $SN_{Au}$ and $SP_{Au}$ .Inset shows the selected area (marked by red circle) 3D surface plot.}
\end{center}
\end{figure}

To further confirm the presence of tent--like morphology, SEM image of sample $SP_{Au}$ has been recorded by tilting the sample with 45 degrees as shown in Figure 3. It is evident from Figure 3 that SiNS of tent shaped is successfully fabricated in sample $SP_{Au}$. The width of SiNS tent structure is of the order of 10$\mu$m marked with red curve in Figure 3 as confirmed from the surface plot reproduced from ImageJ. The surface plot shown in inset of Figure 2 is used to calculate the tent angle by ImageJ software, which is found to be $\sim$ 72 degree for sample $SN_{Ag}$ and 61.570 for
\begin{figure}
\begin{center}
\includegraphics[width=8cm]{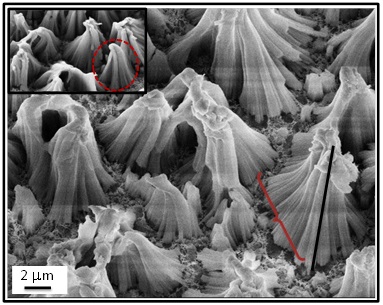}
\caption{SEM image of sample $SP_{Au}$ obtained by tilting the sample. The red and black marked line shows the single tent width and height respectively. Inset shows the bending of Si--tent tip.}
\end{center}
\end{figure} sample $SP_{Au}$. The approximate height of single SiNS tent is 10m in case of sample $SP_{Au}$ marked by black line and the spacing between two consecutive tent is of the order of 1$\mu$m.

It is also evident from the SEM images that the tip of the tent are bent as shown in inset of Figure 3 and looks as if a bunch of silicon nano wires (SiNWs) are clubbed together and bent on one another. The bending of SiNWs on one another form the tent like structure with SiNWs due to that the surface of etched Si wafer texturises in the form of a tent. This kind of texturization can be thought to be originating as an effect of surface tension and can be understood as follows. Synthesis of SiNWs using MIE have been known and is used as a fabrication method as has already been reported[4,26]. The SiNWs thus fabricated may feel the force from the solution and will agglomerate to minimize the surface energy which will be very high if all the wire remain separated making it more unfavourable energetically. The same has been explained by depicting the texturization of tent shaped SiNS in Figure 4. The texturization of c--Si is involved the 4 step here we are showing when the Si wafer and metal nano particles formed ohmic contact[35]. In first step the metal nano particles were deposited on the surface of Si wafer by dipping in metal solution and the metal nano particle deposited Si wafer transfer in the etching solution. After etching the well aligned SiNWs have been formed which is shown in step 2 and the black arrow shows the depth (height) of pores (nanowires) during MIE. As we increase the etching time the length of SiNWs also increases. In step 3 the SiNWs start bending on each other as they are unable to sustain on their own due to increased length. This starts during removal of etched Si wafer from the etching solution the red arrow in circle shows the bending of SiNWs in step 3. In the final step the bunch of SiNWs were bent together and form a tent like structure when the etched Si wafer removed completely from the etching solution. It is very likely that the bunch of SiNWs remain well aligned when they are in the etching solution. The base of SiNWs stay attached to the Si wafer.This way, the SiNWs bend due to the surface tension effect when nanowires are formed by MIE from porosification of metal deposited Si wafer making Ohmic junction. The surface tension shows the elastic tendency of the solution which forces it to acquire the least surface area possible. This can be understood by following simple example.

\begin{figure}
\begin{center}
\includegraphics[width=12cm]{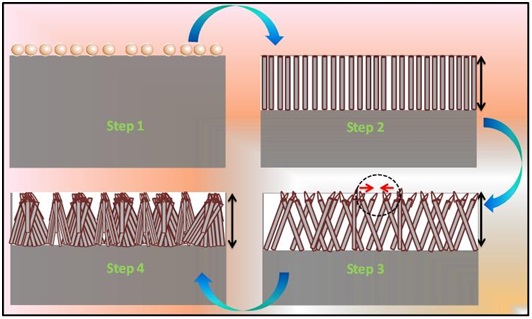}
\caption{Schematic diagram shows the mechanism of texturization of Si wafer in to nano tent of SiNS. Red arrows, in step 3, show the direction of force being felt by the wires as a result of surface tension}
\end{center}
\end{figure}

\begin{figure}
\begin{center}
\includegraphics[width=12cm]{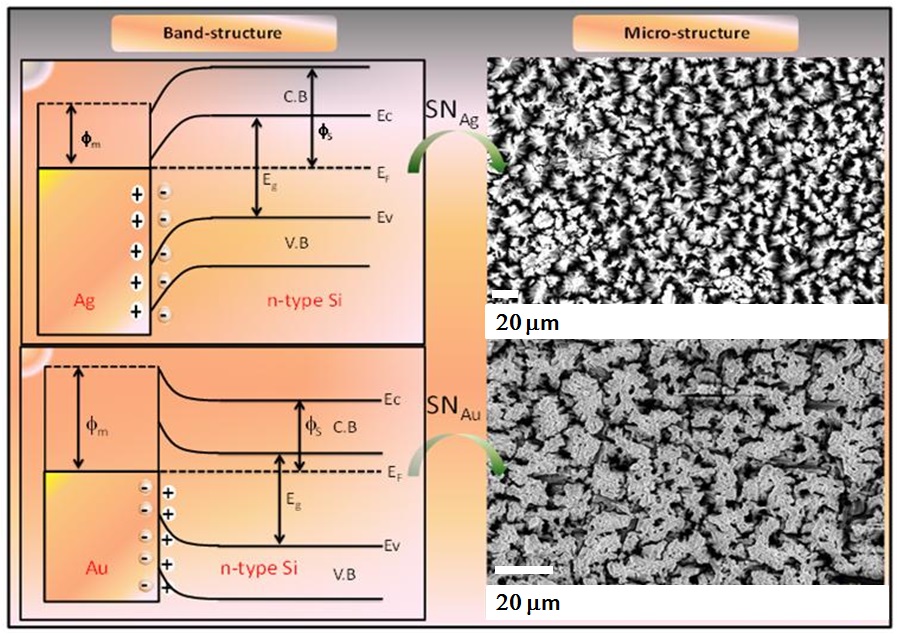}
\caption{Band diagram along with SEM micrographs showing a correlation between band bending and morphology after porosification.}
\end{center}
\end{figure}

It is well known that the surface area of two 1--1 ml drops of water droplet is 0.42 times more than that of one 2 ml drop of water, which shows that the surface area decreases as the SiNWs removed from the etching solution and due to that they bent on one another and form a Si--tent. The same effect has been observed when the cat hairs form cluster after bath. The bunch of wires after bent from their original position form the tent shaped structures which have been depicted in step 4. It is important here to mention here that a clear correlation between the nano--metal/Si junction used for porosification and resultant surface morphology is observed as shown in Figure 5. It is clearly noticeable that the nano--metal.Si combination making Ohmic junction, when etched, results in ten--shaped morphology whereas Schottkey junction nano--metal/Si combination will result in non--tent shaped morphologies. It is also worth mentioning here that the nano tents are very sharp (top angle of $\sim$72 degree) and will be very useful for field emission applications[5].

\section{Conclusions}

In summery, surface morphologies of silver-- and gold-- nanoparticle assisted etching of n-- and p--type wafers have been studied to observe a correlation between obtained nanostructure and nano--metal/silicon (Si) contact. A tent--shaped surface morphology is obtained when an Ohmic nanometal/Si wafer is etched by metal induced etching (MIE). The ten--shaped surface morphologies results irrespective of the wafer type (n-- or p--) as long it forms an Ohmic junction with nano--metal used as catalysis for MIE. The typical height of single Si--tent is $\sim$10 $\mu$ m and the spacing between two consecutive tents is of the order of 1m as estimated using ImageJ. Appropriate control experiments clearly support this correlation. Furthermore, Si wafers making Schottkey contact with nano--metal , used for catalysis, result in non--tent shaped surface morphologies.  It is very likely that surface tension is playing a role in resulting tent--shaped nanostructures of silicon as a result of MIE. These nano--tents with very sharp tips having  angle of 72 degrees appears to be applied for field emission applications.

\subsection*{Acknowledgement} Authors acknowledge Sophisticated Instrumentation Centre (SIC), IIT Indore for SEM measurements. Financial Support from Department of Science and Technology (DST), Govt. of India is also acknowledged. Authors thank MHRD for providing fellowship. Useful discussion with Dr Vivek Kumar (NIT Meghalaya) is also acknowledged

\newpage

\end{document}